\documentclass[a4paper,11pt]{article}
\usepackage{jheppub} 
\usepackage{lmodern}
\usepackage[T1]{fontenc}
\usepackage[dvipsnames]{xcolor}
\hypersetup{
	pdfstartview={FitH},
	pdftitle={Effective Gravitational Couplings of
          Kaluza-Klein Gauge Theories},
	pdfauthor={S. K. Ashok, R. R. John, T. J. Layon, S. Mahato,
          M. Raman},
      }
      \usepackage{parskip}
      \usepackage{cleveref}

      \usepackage{array}

      \usepackage{booktabs}
\DeclareMathOperator{\Tr}{Tr}

\newcommand{\be}{\begin{equation}}
\newcommand{\ee}{\end{equation}}

\newcommand{\one}{{\rm 1\kern -.9mm l}}

\newcommand{\ttA}{ \texttt{A}}
\newcommand{\ttB}{ \texttt{B}}
\newcommand{\ttF}{ \texttt{F}}
\newcommand{\ttH}{ \texttt{H}}

\title{\boldmath Effective Gravitational Couplings of \\
Kaluza-Klein Gauge Theories}

\author[a,d]{Sujay K. Ashok,}
\author[b]{Renjan Rajan John,}
\author[b]{T. J. Layon,}
\author[c,d]{Sujoy Mahato,}
\author[e]{and Madhusudhan Raman}

\affiliation[a]{The Institute of Mathematical Sciences \\
  IV Cross Road, C.I.T. Campus, Taramani, Chennai 600 113, India}
\affiliation[b]{School of Pure and Applied Physics\\
  Mahatma Gandhi University, Kottayam, Kerala 686 560, India}
\affiliation[c]{Harish-Chandra Research Institute\\
  Chhatnag Road, Jhunsi, Allahabad 211 019, India}
\affiliation[d]{Homi Bhabha National Institute\\
  Training School Complex, Anushakti Nagar, Mumbai 400 094, India}
\affiliation[e]{Department of Physics and Astrophysics\\
  University of Delhi, New Delhi 110 007, India\\}

\emailAdd{sashok@imsc.res.in, renjan@mgu.ac.in, layontj@mgu.ac.in\\
  sujoymahato@hri.res.in, mraman@physics.du.ac.in}

\abstract{We study the effective gravitational couplings of
  four-dimensional Kaluza- Klein compactified gauge theories with
  eight supercharges. The class of theories we consider are the pure
  $ \mathrm{SU}(N) $ Yang-Mills theories at admissible Chern-Simons
  levels and the conformal gauge theories with $ 2N $ fundamental
  flavours. The resolvent of the gauge theory plays a crucial role in
  the calculation of these gravitational couplings. The results
  obtained from the Seiberg-Witten geometry are matched against
  independent computations using localisation.}

\begin{document}
\maketitle
\flushbottom

\section{Introduction} 

Gauge theories on four-manifolds have been an active area of
investigation for some decades now, and this body of work represents
perhaps one of the most striking examples of the richness of
\emph{physical mathematics}. Celebrated early efforts
\cite{Witten:1994ev,Moore:1997pc} at outlining a physical approach to
Donaldson invariants have played a significant role, not only in
simplifying many proofs, but also shedding light on the dynamics of
strongly coupled gauge theories.

A widely studied class of theories in this field are the
$ \mathcal{N} = 2 $ supersymmetric gauge theories, whose low-energy
effective dynamics is encoded in the famous Seiberg-Witten geometry
\cite{Seiberg:1994rs,Seiberg:1994aj}. In its simplest avatar --- for
the case of pure $ \mathrm{SU}(2) $ gauge theory --- this geometry is
an elliptic fibration over the $ 1 $-complex dimensional Coulomb
moduli space. More generally, when we speak of the Seiberg-Witten
geometry, we will mean an algebraic curve (whose moduli space is
identified with the \emph{quantum} moduli space of the supersymmetric
gauge theory) and an associated differential. This data is sufficient
to fully (i.e.~nonperturbatively) solve for the low-energy effective
action on the Coulomb branch.

These theories can in fact be defined on arbitrary four-manifolds
$ \mathcal{M} $ using the topological twist \cite{Witten:1988ze}, and
the low-energy effective action includes the gravitational couplings
\cite{Witten:1995gf} to the Euler character and the signature of the
four-manifold:
\begin{equation}
S \supset \frac{1}{32\pi^2} \int _{\mathcal{M}}\left(\operatorname{tr}  R\wedge \star  R\right)  \log \texttt{A}(a_i) + \frac{1}{24\pi^2} \int _{\mathcal{M}} \left(\operatorname{tr} R\wedge R\right) \log \texttt{B}(a_i)~,
\end{equation}
where $R$ is the curvature 2-form and $ a _{i} $ are the vevs of the
adjoint scalar in the $ \mathcal{N}=2 $ vector multiplet. The functions $\ttA$ and $\ttB$ appear 
as measure factors in the $U$-plane integral that computes topological
invariants. Based on arguments leveraging holomorphy, R-symmetry, and
electric-magnetic duality
\cite{Witten:1995gf,Moore:1997pc,Losev:1997tp,Marino:1998bm}, it is
expected that for generic $ \mathcal{N}=2 $ gauge theories, these
gravitational couplings are specified by their Seiberg-Witten geometry
in the following manner:
\begin{equation}
  \label{rels0}
  {\texttt A}(a_i) = \widetilde\alpha \, \text{det} \left(\frac{\mathrm{d} u_i}{\mathrm{d} a_j}\right)^{1/2} \quad \mathrm{and} \quad 
    {\texttt B}(a_i) = \widetilde\beta\,\left[ \Delta_{\text{phys.}}\left( u _{i} \right)\right]^{1/8} \ .
\end{equation}
Here, the $u_i$ are gauge-invariant coordinates on Coulomb moduli
space, and $\Delta_{\text{phys.}}$ is proportional to the discriminant
of the Seiberg-Witten curve. The constants $ \widetilde{\alpha } $ and
$ \widetilde{\beta } $ are not determined at this stage, and are fixed
by an independent computation of the same. This is done by studying
the theory on a solvable background. In \cite{Manschot:2019pog}, a
number of rank-$ 1 $ gauge theories were studied on the
$ \Omega $-deformed $ \mathbb{C}^{2} $, first used in the equivariant
localisation computations of
\cite{Nekrasov:2002qd,Nekrasov:2003rj}. Since the Euler characteristic
and signature of the $ \Omega $ background are quadratic in the
deformation parameters, the gravitational couplings in question can be
read off from a small deformation expansion of the
partition function of the $\Omega$-deformed gauge theory:
\begin{equation}
\epsilon _{1} \epsilon _{2} \log Z = - \ttF + \left( \epsilon _{1}+\epsilon _{2} \right) \ttH + \epsilon _{1}\epsilon _{2} \log \texttt{A} + \frac{\epsilon _{1}^{2}+\epsilon _{2}^{2}}{3} \log \texttt{B} + \cdots \ .
\end{equation}
In this way, by comparing the results of curve and localisation
computations, they were able to determine the dependence of the
constants $ \widetilde{\alpha } $ and $ \widetilde{\beta } $ on
hypermultiplet masses and the strong-coupling scale. Subsequent work
\cite{John:2022yql} generalised their analysis of mass-deformed
$ \mathcal{N}=4 $ theories to all higher-rank gauge theories with
unitary gauge groups, and  \cite{Closset:2021lhd} studied the
gravitational couplings of rank-$ 1 $ Kaluza-Klein theories using
similar techniques. 

In this paper we will study the couplings $ \texttt{A} $ and
$ \texttt{B} $ for higher-rank four-dimensional $ \mathcal{N}=2 $
theories that are arrived at via Kaluza-Klein compactification of
$ \mathcal{N}=1 $ theories in five dimensions. Working with
higher-rank gauge groups presents certain obvious challenges. For
example, computing both gravitational couplined using \cref{rels0}
requires knowledge of the relation between the gauge-invariant moduli
$ u _{i} $ and the classical vacuum expectation values $ a _{i} $,
which becomes increasingly complicated as we go to higher orders in
the instanton expansion. In order to circumvent this difficulty, we
introduce a two-step procedure.  First, we use equivariant
localisation to compute $ 1 $-point functions of gauge-invariant
chiral observables, as in \cite{Fucito:2011pn, Billo:2012st,
  Ashok:2016ewb}. The generating function of these $ 1 $-point
functions is the resolvent of the gauge theory, and it is completely
determined by the data of the Seiberg-Witten curve
\cite{Cachazo:2002ry, Ashok:2017bld}. We show that composing these two
operations gives us the sought after $u _{i}$ vs.~$a _{j}$ relations
as a power series in the instanton counting parameter. This makes it
possible to compute the $ u _{i} (a _{j}) $ without having to compute
and invert period integrals.

These results are then compared with the couplings $ \texttt{A} $ and
$ \texttt{B} $ read off from a small deformation expansion of the
$\Omega$-deformed partition function. Requiring that the two
independently computed quantities agree fixes the constants
$ \widetilde{\alpha } $ and $ \widetilde{\beta } $.

We have focused on the case of pure $ \mathrm{SU}(N) $ theories and on
the case of conformal theories with $ N _{f} = 2N $ fundamental
hypermultiplets in five dimenions, both Kaluza-Klein compactified on
$ \mathbb{S}^{1} $. In the former case, we find
perfect agreement between curve and localisation computations for all
admissible Chern-Simons levels once the logarithmic terms arising from
the perturbative sector of the theory are accounted for. The latter
case is perhaps more interesting in that in addition to perturbative
contributions, we must also account for nondynamical discrepancies
between the results of equivariant localisation and those coming from
the Seiberg-Witten geometry.  Since we are interested in gravitational
couplings, on dimensional grounds these constants can
only be functions of $ q $, the instanton counting parameter.  In line
with prior work on conformal gauge theories \cite{John:2022yql}, we
show that all $ k $-instanton discrepancies can be attributed to the
so-called $ \mathrm{U}(1) $ factor. As we show, this follows as a 
consequence of a specific choice in the parametrisation of the 
gauge invariant coordinates on the Coulomb moduli space.

This paper is organised as follows. In Section \ref{SWofKK} we review
the Seiberg-Witten geometry of Kaluza-Klein compactified SU$(N)$ gauge
theories with eight supercharges. In Section \ref{LowRank5d}, we list
our results for the gravitational couplings for low-rank gauge groups.
We generalize to the case with $N_f=2N$ fundamental flavours in
Section \ref{AddingFlavour} and discuss the mismatch of the curve and
localisation results in this case in Section \ref{u1factor}. We also
analyse the four-dimensional limit of our results in this section. We
summarize and discuss our results in Section \ref{Discussion}. A brief
review of localisation techniques is presented in
\Cref{localizationresults} and a discussion of the perturbative
contributions can be found in \Cref{Zpert}.

\acknowledgments

We would like to thank Marco Bill\`o, Dileep Jatkar, Alberto Lerda, and
Tanmoy Sengupta for helpful discussions. SA is grateful for the
hospitality of the theory group at the Dipartimento di Fisica,
Universit\`a di Torino, Italy during the completion of this work. MR
is supported by a DST-INSPIRE Faculty Fellowship.

\section{Seiberg-Witten Geometry of Kaluza-Klein Theories }
\label{SWofKK}

In this section we review the Seiberg-Witten geometry of five dimensional 
supersymmetric gauge theories with $ \mathrm{SU}(N) $ gauge group,
Kaluza-Klein compactified on $\mathbb{S}^1$.

\subsection{The Curve and the Differential}

The Seiberg-Witten curve for these theories was first studied in the
work of \cite{Nekrasov:1996cz} (see also \cite{Brini:2009nbd} where
the gauge theory is geometrically engineered) and takes the form:
\begin{equation}
\label{SWv2}
Y^2 =P_N(Z)^2 - 4(\beta\Lambda)^{2N} Z^{N-\kappa} \ ,
\end{equation}
where 
\begin{equation}
\label{5dpureSWcurve}
P_N(Z) = Z^N + \sum_{i=1}^{N-1}(-1)^i Z^{N-i}U_i + (-1)^N ~.
\end{equation}
Here $\beta = 2\pi R$ is the circumference of the $\mathbb{S}^1$, the
parameters $U_i$ are gauge-invariant coordinates on the Coulomb branch
of the Kaluza-Klein reduced five-dimensional theory, and $\kappa$ is
the coefficient of the five-dimensional Chern-Simons term, which takes
integral values $\left\vert \kappa \right\vert \le N-1$
\cite{Intriligator:1997pq}. We will henceforth refer to any $ \kappa $
satisfying this condition as an \emph{admissible} Chern-Simons
level. 

In the classical limit, the Coulomb moduli reduce to elementary
symmetric polynomials in the $A_u$, the classical vevs of the
Kaluza-Klein theory:
\begin{equation}
U_i^{(\mathrm{c})} = \sum_{1 \leq u_1 < \cdots < u_i \leq N}^N A_{u_1} \ldots A_{u_i}~.
\end{equation}
The $A_u$ are related to the four-dimensional vevs $a_u$ of the
adjoint scalar $ \Phi $ by the exponential map
$A_u = e^{\beta a_u}$.\footnote{Similarly if we use the relation
  $Z= e^{\beta z}$, then in the $\beta\rightarrow 0$ limit we recover
  the Seiberg-Witten curve of the pure ${\cal N}=2$ gauge theory in
  four dimensions if we appropriately scale the $Y$ coordinate.}
For the SU$(N)$ theory, as is evident from \cref{5dpureSWcurve}, we
have set $U_N = \prod_{u=1}^N A_u = 1$.  This ensures that in the
classical limit, the quantum gauge polynomial $P_N(Z)$ reduces to the
classical one:
\begin{equation}
P_N^{(\mathrm{c})}(Z) = \prod_{u=1}^N(Z-A_u)~.
\end{equation}

The associated Seiberg-Witten differential is given by
\cite{Brini:2009nbd}:
\begin{equation}
  \label{eq:sw-differential}
\lambda_{\mathrm{SW}} = \Psi(Z) \frac{\mathrm{d} Z}{Z}~,
\end{equation}
where
\begin{equation}
  \label{eq:psi-z}
\Psi(Z) = \log\left( Z^{-\frac{N}{2}} \frac{(P_N(Z) + Y)}{2} \right)~.
\end{equation}

\subsection{The Resolvent} 
\label{sec:resolvent}
We have just seen that classically, the $U_i$ that appear in the
Seiberg-Witten curve are elementary symmetric polynomials of the
Coulomb vevs $A_u$. In order to make contact with the gravitational
couplings that are the object of our study, we will need to compute
instanton corrections to these Coulomb moduli for admissible values of
the Chern-Simons coupling. We do this in two steps.

First, we compute the resolvent of the gauge theory, which will
provide a precise relationship between Coulomb moduli $U_i$ and the
one point function of the chiral observables of the
gauge theory. Next, we calculate the chiral observables directly 
using localisation methods. By composing these two steps, we can
compute the instanton corrections to the $U_i$ at any point on the
Coulomb moduli space, order by order in the instanton expansion.  We
pause to mention that while the resolvent is typically used to study
how the trace relations are modified by quantum corrections
\cite{Cachazo:2002ry}, our goal will be to obtain the
$U_i$ vs.~$V_{\ell}$ relations for $i<N$.

Recall that the resolvent of the gauge theory is the generating
function of the chiral correlators. For the five-dimensional gauge
theory, it is defined to be \cite{Wijnholt:2004rg, Ashok:2017bld}:
\begin{equation}
T(Z) := \left\langle \Tr \coth\frac{\beta(z-\Phi)}{2 }\right\rangle = \frac{2}{\beta}\frac{\mathrm{d} }{\mathrm{d} z}\left\langle \Tr\log \left( 2\sinh\frac{\beta(z-\Phi)}{2 } \right)\right\rangle~.
\end{equation}
If we expand this for large $Z$, we obtain
\begin{equation}
\label{resolventdefinition}
T(Z)= N + 2\sum_{\ell} \frac{V_{\ell}}{Z^{\ell}}~,
\end{equation}
where  the $V_{\ell}$ are the one-point functions:
\begin{equation}
V_{\ell} = \left\langle \Tr~ e^{\beta\ell \Phi} \right\rangle~.
\end{equation}
The resolvent is directly given in terms of the Seiberg-Witten
differential associated to the curve \cite{Ashok:2017bld} and is given
by:
\begin{equation}
T(Z) = 2\, Z\,\frac{\mathrm{d} \Psi(Z)}{\mathrm{d} Z}~.
\end{equation}
Substituting for $\Psi(Z)$ from \cref{eq:psi-z}, we find that
\begin{equation}
\begin{aligned}
T(Z) &= 2\, Z \frac{\mathrm{d} }{\mathrm{d} Z} \log\left(Z^{-\frac{N}{2}} \frac{P_N(Z) + Y}{2} \right) \ , \\
 & = 2\,Z \,\frac{P_N'(Z)}{Y} +(N- \kappa)\left(1-\frac{P_N(Z)}{Y} \right) -N~.
\end{aligned}
\end{equation}
By taking the $\beta\rightarrow 0$ limit, one can check that this is
the five-dimensional uplift of the resolvent of the four-dimensional
pure ${\cal N}=2$ gauge theory that was computed in
\cite{Cachazo:2002ry} using the Konishi anomaly.

Expanding this expression for large $Z$ and comparing with
\cref{resolventdefinition} allows us to read off relations expressing
chiral correlators in terms of sums of products of Coulomb moduli,
which are easy to invert. At $ 1 $-instanton, we have the following
relations \cite{Ashok:2017bld}:
\begin{equation}
\begin{aligned}
     U_1(\kappa)&=V_1-(\beta\Lambda)^{2N}\delta_{\kappa,1-N}\ , \\
     U_2(\kappa)&=\frac{1}{2}\left(V_1^2-V_2\right)+(\beta\Lambda)^{2N}\left(V_1\delta_{\kappa,1-N}+\delta_{\kappa,2-N}\right)~,
\end{aligned}
\end{equation}
and so on. We find the universal relation at $ 1 $-instanton for all
values of $n$: \footnotesize
\begin{equation}
\begin{aligned}
  U _{n} \left( \kappa   \right) &= \left( -1 \right)^{n} \sum_{\left\lbrace m _{p}  \right\rbrace \vdash n }^{} \prod _{p = 1}^{n} \frac{\left( -V _{p} \right)^{m _{p}}}{m _{p}! p ^{m _{p}}} + \left( -1 \right)^{n} \left( \beta \Lambda  \right)^{2N} \left[ \sum_{\ell = 0 }^{n-1} \left( \sum_{\left\lbrace m _{s} \right\rbrace \vdash \ell }^{} \prod _{s=1}^{\ell } \frac{V _{s}^{m _{s}}}{m _{s}! s ^{m _{s}}} \right) \delta _{\kappa  , n - N - \ell } \right] \ ,
\end{aligned}
\end{equation}
\normalsize
where the sum over $ \left\lbrace m _{p} \right\rbrace \vdash n $
indicates a sum over partitions of $ n $ such that
$ \sum_{p}^{} p \, m _{p} = n $. We emphasise that although the above
result is only true at $ 1 $-instanton, it is possible to compute $ k
$-instanton corrections to the same in a straightforward manner. 

We see that for $\kappa>0$, the relation between the $U _{i}$ and the
$V _{\ell }$ are exactly what one would expect from the classical
theory; the above relation exchanges power sums for elementary
symmetric polynomials. However, for $\kappa <0$, we see that these
relations are corrected by instantons.

The next step is to compute chiral correlators using supersymmetric
localisation. We refer the reader to \Cref{localizationresults} for
details and simply present the $ 1 $-instanton result for chiral
correlators in the Kaluza-Klein reduced pure SU$(N)$ gauge theory
\cite{Ashok:2017bld}:
 \begin{equation}
 \label{Velllocmaintext}
    V_{\ell }= \sum_{u=1}^N  A_u^{\ell } +\ell ^2(\beta\Lambda)^{2N}\sum_{u=1}^N\left[\frac{A_u^{N-2+\ell -\kappa}}{\prod_{u\ne v}(A_u-A_v)^2}\right]+\mathcal O\left((\beta\Lambda)^{4N}\right) \ .
 \end{equation}
 where we have imposed the $\mathrm{SU}(N)$ condition
 $\prod_{u=1}^N A_u =1$.

\subsection{Proposed Formulas for $ \texttt{A} $ and $ \texttt{B} $}
\label{sec:proposed-formulas-}
As we have reviewed in the Introduction, the proposal for the
$\texttt{A}$ and $\texttt{B}$ functions that appear as measure factors
in the $U$-plane integral are
\begin{equation}
\label{rels1}
   {\texttt A} = \widetilde\alpha~ \text{det} \left(A_j\frac{\mathrm{d} U_i}{\mathrm{d} A_j}\right)^{1/2} \quad \mathrm{and} \quad  {\texttt B} = \widetilde\beta~ \left[\Delta_{\text{phys.}}\left( U _{i} \right)\right]^{1/8}~.
\end{equation}
Here, $\Delta_{\text{phys.}}$ is called the ``physical discriminant''
and is proportional to the mathematical discriminant of the
Seiberg-Witten curve, which in turn is a polynomial in the Coulomb
moduli. Given that we have an algorithmic way to compute the $U_i$ as
a function of the Coulomb vevs, we are now in a position to compute
these two functions as a power series in the instanton counting
parameter $q=(-1)^N\, (\beta\Lambda)^{2N}$.

We check these results by explicitly computing these functions in the
$\Omega$-background.  The instanton contributions are obtained by
doing the contour integrals of the (five-dimensional uplift of the)
Nekrasov integrand. The integrands, the contours, and the
$ 1 $-instanton results for all SU$(N)$ gauge groups are summarised in
Appendix \ref{localizationresults}. There is, in addition to this, a
leading term that survives the $q\rightarrow 0$ limit. This is the
perturbative contribution and is computed for the $\Omega$-background
in the five-dimensional case in \cite{Nakajima:2005fg}. We review this
in detail in Appendix \ref{Zpert}. In the sections that follow, the
sum of these two contributions is referred to as the localisation
contribution.

\section{Results for Low Rank Gauge Theories}
\label{LowRank5d}

In this section we present results for low-rank gauge theories, by way
of illustration.

\subsection{SU(2)}
The Seiberg-Witten curve for SU$(2)$ theory with Chern-Simons level
$\kappa$ takes the form
\begin{equation}
Y^2 = (Z^2 - Z U+1)^2 - 4q Z^{2-\kappa} \ .
\end{equation}
Here, we have already identified $q=(\beta\Lambda)^{4}$ with the
instanton counting parameter, in order to facilitate the match with
the localisation results easier. (We have also set $ U _{1} \equiv U $
in order to lighten the notation.) The discriminant of the curves for
various $ \kappa $ are easily computed to be
\begin{equation}
  \begin{aligned}
\Delta_{\kappa=0}&= 256q^2\left(U^4 -8(1+q)U^2+16(1-q)^2\right) \ , \\
\Delta_{\kappa=\pm1}&=256q^2 \left(U^4+qU^3-8U^2-36qU-27q^2+16\right)~.
  \end{aligned}
\end{equation}
We then read off the physical discriminant as
\begin{align}
\Delta^{(\kappa)}_{\text{phys.}}=\frac{\Delta^{(\kappa)}}{256q^2} \ .
\end{align}

Recall that the relation between $U$ and $V$ is given via the
resolvent as reviewed in \Cref{sec:resolvent}. For the SU$(2)$ theory
with $\kappa=0$, this relation is simply $U(\kappa)=V$. The chiral
correlator $V$, in turn can be computed using localisation
techniques. Putting all this together we have (with
$A \equiv A _{1} = A _{2}^{-1}$):
\begin{equation}
  \label{ccfromsu2loc}
  \begin{aligned}
     U&=A+A ^{-1}+\frac{A+A ^{-1}}{\left(A-A ^{-1}\right)^2}q+5\frac{A+A ^{-1}}{\left(A-A ^{-1}\right)^6}q^2+\cdots  \quad\text{for}\quad \kappa=0\ , \\
         U&=A+A ^{-1}+\frac{2}{\left(A -A ^{-1}\right)^2}q+5\frac{(A +A ^{-1})}{\left(A-A ^{-1}\right)^6}q^2+\cdots \quad\text{for}\quad \kappa=\pm 1 \ .
  \end{aligned}
 \end{equation}
 Given these, one can evaluate $\log \texttt{A}$ and $\log \texttt{B}$
 using the proposed formulae in \cref{rels1}. For $\kappa=0$, we obtain
\begin{equation}
  \begin{aligned}
\log \ttA &= \frac12 \log\left(A -A ^{-1}\right) -\frac{q}{2}\frac{A^2\left(A^4+6A^2+1\right)}{\left(1-A^2\right)^4}\\
&\hspace{3.3cm}-q^2\frac{A^4\left(A^8+62A^6+178A^4+62A^2+1\right)}{4\left(1-A^2\right)^8}+\cdots \ , \\
\log \ttB & =\frac12\log\left(A -A ^{-1}\right)-\frac{q}{2}\frac{A^2\left(A^4+10A^2+1\right)}{\left(1-A^2\right)^4}\\
&\hspace{3.3cm}-\frac{q^2}{4}\frac{A^4\left(A^8+94A^6+314A^4+94A^2+1\right)}{\left(1-A^2\right)^8}+\cdots \ .
  \end{aligned}
\end{equation}
For $\kappa = \pm 1$ we have
\begin{equation}
\begin{aligned}
\log \ttA &=\frac12 \log\left(A-A ^{-1}\right) -2q\frac{A^3\left(1+A^2\right)}{\left(1-A^2\right)^4}-\frac{q^2}{2}\frac{A^6\left(33+86A^2+33A^4\right)}{\left(1-A^2\right)^8}+\cdots \ , \\
\log \ttB &=\frac12\log \left(A -A ^{-1}\right) +\frac{q}{8}\frac{A \left(1-25A^2\left(1+A^2\right)+A^6\right)}{\left(1-A^2\right)^4} \\
&\quad \quad \quad - q^2\frac{A^2\left(1-8A^2+415A^4+1200A^6+415A^{8}-8A^{10}+A^{12}\right)}{16\left(1-A^2\right)^8}+\cdots \ .
\end{aligned}
\end{equation}
Comparing with the localisation results we find a perfect
match with the perturbative and instanton contributions provided we set
 \begin{equation}
  \widetilde\alpha = (\beta\Lambda)^{-\frac12} \quad\text{and}\quad \widetilde\beta = (\beta\Lambda)^{-\frac12}~.
\end{equation}
 This is our first check of
\cref{rels1}.\footnote{In \cite{Closset:2021lhd} the check of the
  formulae was done for $\kappa=0$, as an expansion for large $A$. In
  that reference, the $U$ vs. $A$ relation was obtained by calculating
  the period integral of the Seiberg-Witten differential.}

\subsection{SU(3)}

The Seiberg-Witten curve for the SU$(3)$ gauge theory takes the form
\begin{equation}
Y^2 = (Z^3 - Z^2 U_2 + Z U_1-1)^2 - 4q Z^{3-\kappa}~.
\end{equation}
In this case we have identified $q=(\beta\Lambda)^6$ with the
instanton counting parameter. The resulting expressions are quite
lengthy to list out for all cases, so in what follows we present the
main results only for the $\kappa=0$ case. The discriminant of the
curve in this case is given by
\begin{equation}
\begin{aligned}
  \Delta &= 4096q^3\Big[16(U_1^6+U_2^6)+U_1^4 U_2^4 -8 U_1^2 U_2^2 \left(U_1^3+U_2^3\right) + (68-q) U_1^3 U_2^3 \\
  &\quad +36  (q-4) U_1 U_2 \left(U_1^3+U_2^3\right) -27 (19 q-10) U_1^2 U_2^2 +729 (1-q)^3 \\
  &\quad \quad +27 (q (20-q)+8) \left(U_1^3+U_2^3\right) + 243 \left(5
    q^2-q-4\right) U_1 U_2\Big]~.
\end{aligned}
\end{equation}
The physical discriminant is given by
\begin{equation}
\Delta_{\text{phys.}}=\frac{\Delta}{4096q^3}~.
\end{equation}

As before, the resolvent tells us how the $U_k$'s are related to the
chiral correlators $V_{\ell}$ and we find
\begin{equation}
\begin{aligned}
U_1 &= \sum_{u=1}^3A_u-2q\frac{A_1^2A_2^2+A_2^2A_3^2+A_1^2A_3^2 - A_1-A_2-A_3}{(A_1-A_2)^2(A_2-A_3)^2(A_3-A_1)^2} + \cdots\ , \\
U_2 &=  \sum_{u<v} A_uA_v -2q\frac{A_1^2+A_2^2+A_3^2-A_1A_2-A_2A_3-A_3A_1}{(A_1-A_2)^2(A_2-A_3)^2(A_3-A_1)^2} + \cdots \ .
\end{aligned}
\end{equation}
Of course, in the above expression, we have
$ A _{3} = \left( A _{1} A _{2}\right)^{-1} $. We then find
\footnotesize
\begin{equation}
\begin{aligned}
\log \ttA &= \frac12 \log \operatorname{det} A _{j}\frac{\mathrm{d} U_i}{\mathrm{d} A_j}  = \frac12 \log\left[\frac{(1-A_1^2 A_2)(1-A_1 A_2^2) (A_1-A_2)}{A_1^2 A_2^2}\right]\\
&\quad +\frac{2q A_1^3 A_2^3}{(A_1-A_2)^4~(1-A_1^2 A_2)^4~(1-A_1 A_2^2)^4}\Big[\left(A_2^9+A_2^6\right) A_1^{10}+\left(A_2^{10} -7 A_2^7+A_2^4\right) A_1^9 \\
&\quad \quad -7 A_2^3 \left(A_2^3-1\right){}^2 A_1^7 +\left(A_2^{10}+14 A_2^7+14 A_2^4+A_2\right) A_1^6-54 A_2^5 A_1^5 -7 A_2 \left(A_2^3-1\right){}^2 A_1^3\\
&\quad \quad \quad +\left(A_2^9+14 A_2^6+14 A_2^3+1\right) A_1^4+\left(A_2^6-7 A_2^3+1\right) A_1+A_2^4+A_2\Big] + \cdots ~.
\end{aligned}
\end{equation}
\begin{equation}
\begin{aligned}
\log \ttB &=\frac18 \log \Delta_{\text{phys.}} = \frac12\log \left[\frac{\left(1-A_1 A_2^2\right) \left(A_1^2 A_2-1\right) \left(A_1-A_2\right)}{A_1^2 A_2^2}\right]\\
&\quad -\frac{qA_1^2A_2^2}{8\left(A_1-A_2\right){}^4 \left(A_1^2 A_2-1\right){}^4 \left(A_1 A_2^2-1\right){}^4}\Big[ 
A_2^9 A_1^{12}-25 \left(A_2^{10}+A_2^7\right) A_1^{11}\\
&\quad \quad -5 (A_2^5 A_1^{10} +A_2A_1^2)\left(5 A_2^6-32 A_2^3+5\right) +\left(A_2^{12}+18 A_2^9+18 A_2^6+A_2^3\right) A_1^9 \\ 
&\quad \quad \quad +(A_2^4A_1^8 +A_2^2 A_1^4)\left(160 A_2^6-353 A_2^3+160\right) -A_2^2 \left(25 A_2^9+353 A_2^6+353 A_2^3+25\right)A_1^7\\
&\quad \quad \quad \quad +6 A_2^3 \left(3 A_2^6+224 A_2^3+3\right) A_1^6-A_2 \left(25 A_2^9+353 A_2^6+353 A_2^3+25\right) A_1^5\\
&\quad \quad \quad \quad \quad +\left(A_2^9+18 A_2^6+18 A_2^3+1\right) A_1^3-25 \left(A_2^5+A_2^2\right) A_1+A_2^3\Big] + \cdots ~.
\end{aligned}
\end{equation}
\normalsize We have verified that these results lead to a precise match of the results
for the $\ttA$ and $\ttB$ functions coming from localisation at the
$ 2 $-instanton level for all admissible Chern-Simons levels if we
set
\begin{equation}
\widetilde{\alpha } = \left( \beta \Lambda  \right)^{-\frac{N(N-1)}{4} } \quad \mathrm{and} \quad \widetilde{\beta } = \left( \beta \Lambda  \right)^{-\frac{N(N-1)}{4} } \ ,
\end{equation}
for $ N=3 $. For higher-rank gauge groups, the above result can be
established in perturbation theory by expanding the $ 1 $-loop
contribution to the $ \Omega $-deformed partition function to
quadratic order in the deformation parameters and comparing logarithmic
terms.

\section{Adding Fundamental Flavours}
\label{AddingFlavour}

We now extend our analysis of the previous sections to gauge theories
with fundamental flavour. In particular we restrict ourselves to the
SU$(N)$ gauge theory, set the Chern-Simons level $\kappa=0$, and
consider the asymptotically conformal case in which we have $N_f=2N$
fundamental flavours.

\subsection{The Curve and the Differential} 
 
The Seiberg-Witten curve of the gauge theory \cite{Nekrasov:1996cz} is
given by
\begin{equation}
  \label{eq:sw-curve-superconformal}
y^2 = P_N^2(Z) -g^2\, B_{N_f}(Z)~,
\end{equation}
where the relevant functions are given by:
\begin{equation}
\begin{aligned}
  P_N(Z) &= Z^{-\frac{N}{2}}\left(Z^N - \tilde{U}_{1}\, Z^{N-1} + \cdots + (-1)^{N-1} \tilde{U}_{N-1} \, Z + (-1)^N \right)~,\\
B_{N_f}(Z) &= Z^{-\frac{N_f}{2}}\left(Z^{N_f} -S_1 Z^{N_f-1} +\cdots +(-1)^{N-1}S_{N_f-1}  + (-1)^{N_f} \right)~.
  \label{flavourP}
  \end{aligned}
\end{equation}
Here, the $ \tilde{U}_{i} $ are proportional to the Coulomb moduli
$ U _{i}$ up to the addition of some constants on the Coulomb
branch.\footnote{This is analogous to the relationship between the
  moduli $ u $ and $ \tilde{u} $ used in \cite{Manschot:2019pog} for
  the case of the $ N _{f} = 4 $ theory. The two moduli agree up to
  constants on the Coulomb branch.} Our choice of parametrisation is
\begin{equation}
\tilde{U}_{i} = \left(\frac{1-q}{1+q}\right) \, U _{i} \ .
\end{equation}
A more extensive discussion of the rationale behind this prefactor
will be given in the subsequent sections.

The $S _{k}$ in \cref{flavourP} are the Weyl-invariant symmetric
polynomials in the exponentiated masses
{$M_{\ell} = e^{\beta m_{\ell}}$} of the fundamental flavours, with
$m_{\ell}$ being the four-dimensional masses; the flavour function can
be equivalently written as
\begin{equation}
B_{N_f} = Z^{-\frac{N_f}{2}}\prod_{i={\ell}}^{N_f} (Z-M_{\ell})~,
\end{equation}
As is clear from \cref{flavourP}, we will restrict ourselves to the case in which the flavour symmetry is
SU$(2N)$, by setting $\prod_{\ell=1}^{2N} M_{\ell} =1$. It is easy to
check that in the four dimensional limit $\beta\rightarrow 0$, the
Seiberg-Witten curve in \cref{eq:sw-curve-superconformal} reduces to
the well-known curve for the four-dimensional theory at the
$\beta^{2N}$ order.

In order to identify the results obtained from the Seiberg-Witten
geometry with those obtained via localisation, it is important to
relate the dimensionless parameter $g^2$ with the instanton counting
parameter $q$. This can be achieved by realizing the gauge theory as a
system of two parallel NS$5$ branes transverse to $ N $ parallel D$4$
branes and lifting the brane configuration to M-theory
\cite{Witten:1997sc, Gaiotto:2009we, Bao:2011rc, Bao:2013pwa}. The ratio of the asymptotic positions
of the NS5 branes is identified with the Nekrasov counting
parameter. Matching the two curves in this way leads to the relation
\begin{equation}
\label{gvsq}
g^2=\frac{4q}{(1+q)^2}~. 
\end{equation}

The Seiberg-Witten differential associated to the curve takes the 
form:
\begin{equation}
\label{Psiforconformal}
\Psi(Z) = \log\left( \frac{(1+q)}{2}\, \frac{P_N(Z)+y}{\mu^N}\right)~.
\end{equation}

\subsection{The Resolvent and the Proposal Revisited}

The next step is to express the $U_k$ that appear in the curve in
terms of the chiral correlators, which can be computed using
localisation methods. The relation between the two is once again given
in terms of the resolvent of the gauge theory. As in the case without
flavour, the generating function of chiral correlators is obtained
from $\Psi(Z)$ by the relation
\begin{equation}
\label{resolventconformal}
T(Z) = 2\,Z\,\frac{\mathrm{d} \Psi(Z)}{\mathrm{d} Z}~.
\end{equation}
The factor of $\frac{1+q}{2}$ in
\cref{Psiforconformal} ensures that in the large-$Z$ expansion of the
resolvent has the correct form $T(Z) = N + \cdots$.

Recall that the resolvent has a large-$Z$ expansion in terms of
$V_{\ell}= \left\langle \Tr e^{\beta\ell \Phi} \right\rangle$ as shown
in \cref{resolventdefinition}.  In order to more easily deal with the
large-$Z$ expansion that we need, we rewrite $\Psi(Z)$ as follows:
\begin{equation}
\label{SWdifferential}
  \begin{aligned}
\Psi(Z) &= \frac{1}{2}\log\left(\frac{(1+q)^2}{4\mu^{2N}} \left(P_N(Z)+y\right)^2\right)\ , \\
&=  \frac{1}{2}\log\left(\frac{(1+q)^2}{4\mu^{2N}}\left(P_N^2(Z) - y^2\right) \frac{\left(P_N(Z)+y\right)}{\left(P_N(Z)-y\right)} \right)\ , \\
&= \frac12\log\left(\frac{qB_{N_f}(Z)}{\mu^{2N}}\right) + \frac12\log\left(\frac{P_N(Z)+y}{P_N(Z)-y} \right)~.
\end{aligned}
\end{equation}
This is essentially the five-dimensional uplift of the analysis that
was done for the four-dimensional ${\cal N}=2$ gauge theory with
$N_f=2N$ fundamental flavours in \cite{Billo:2012st,Ashok:2019rwa}.
Substituting this into \cref{resolventconformal}, and expanding the
resulting expression for large $Z$, we express the $V_{\ell}$ in terms
of the $U_k$. Inverting these relations, we find the following
relations between Coulomb moduli and chiral correlators for the
$ \mathrm{SU}(N) $ gauge theory with $N_f=2N$ flavours:
\begin{align}
\label{ccrelationU1}
U_1 & =  V_1 + \frac{qS_1}{1-q} \ , \\
\label{ccrelationU2}
U_2&= -\frac12V_2 + \frac{(1+q)}{2(1-q)} V_1^2 - \frac{q}{1-q}V_1S_1+\frac{q}{1-q}S_2 ~.
\end{align}
and so on. Once the $V_{\ell}$ are computed --- here, via localisation
--- we have all the ingredients needed to compute the gravitational
couplings from the Seiberg-Witten geometry of the gauge theory. We
once again use the relations:
\begin{equation}
\label{relsv2}
   {\texttt A} = \widetilde\alpha~ \text{det} \left(A_j\frac{\mathrm{d} U_i}{\mathrm{d} A_j}\right)^{1/2} \quad \mathrm{and} \quad  {\texttt B} = \widetilde\beta~ \left[\Delta_{\text{phys.}}\left( U _{i} \right)\right]^{1/8}~.
\end{equation}
While one could, for example, just as well imagine using the
$ \tilde{U}_{i} $ instead of the $ U _{i} $ in the above relations, we
will argue that the above choice is the most appropriate one in the
following sections. To compute $\log\ttB$, one needs to relate the
discriminant of the Seiberg-Witten curve and the physical
discriminant. We define
\begin{equation}
\label{curvevsphysical}
\Delta_{\mathrm{phys.}}(U_i) = \frac{\Delta_{\mathrm{SW}}(U_i) }{\mathrm{Coeff}_{U _{N-1}^{4N}}\Delta _{\mathrm{SW}}}~,
\end{equation} 
where ${4N}$ is the highest degree of $U_{N-1}$ that appears in the
discriminant of the Seiberg-Witten curve.

\subsection{Low Rank Results}
In this section, we summarise our findings for gauge groups with low
ranks. Since the resulting expressions in all cases match perfectly
\emph{up to} purely $ q $-dependent (and therefore non-dynamical)
factors, our presentation will largely focus on these
discrepancies. 

\subsubsection{SU(2)}

The Seiberg-Witten curve for this case is given by
\begin{equation}
Y^2 = \left(Z^2- \tilde{U} Z+1\right)^2 - \frac{4q}{(1+q)^2}\prod_{\ell=1}^4(Z-M_\ell)~.
\end{equation}
The parameter $U$ is determined in terms of the relation
\cref{ccrelationU1}, with the chiral correlator $V$, that can be
determined from localisation to be
\begin{equation}
  V = A + A ^{-1} + \frac{q}{A \left(A^2-1\right)^2}\left[ A^2 \prod
    _{\ell=1} ^{4} \left( A - M _{\ell} \right) + \prod _{\ell=1} ^{4} \left( A
      M _{\ell}-1 \right) \right] + O(q^2) \ .
\end{equation}
Proceeding as before, we first compute $U$ using \cref{ccrelationU1}, and then from \cref{relsv2}, we obtain  
\begin{equation}
\begin{aligned}
\log\ttA &= \frac12 \log\left(A-A ^{-1}\right)+\frac{q}{2(A^2-1)^4}\bigg[ 1- \left(A^6+A^2\right) \left(S_2+6\right)\\
& \hspace{2.3cm}-6 A^4 \left(S_2+1\right)+4 \left(A^5+A^3\right) \left(S_1+S_3\right)+A^8\bigg] + \cdots ~.\\
\log\ttB &= \frac12\log\left(A-A ^{-1} \right) + \frac18 \log\left(\prod_{\ell=1}^4(A-M_\ell)(A^{-1}-M_\ell)  \right) \\
&\hspace{2.3cm} -\frac{q}{8\left(A^2-1\right)^4}\Big[12+(S_1+S_3)(A -25A^3-25A^5+A^7)\\
&\hspace{3cm} \quad + 8A^4(19+5S_2)+4(S_2-10)(A^2+A^6)  +12 A^8\Big] + \cdots \ .
\end{aligned}
\end{equation}
As we have mentioned, on comparing these results --- obtained from the
Seiberg-Witten geometry --- with a localisation computation of the
same, one finds a non-dynamical (i.e.~independent of Coulomb vevs),
purely $ q $-dependent mismatch that has interesting parallels with
the results found for other asymptotically conformal models in the
four-dimensional cases studied in
\cite{Manschot:2019pog,John:2022yql}. We find that up to
$ 2 $-instantons,
\begin{equation}
  \label{eq:discrepancy-su2}
\begin{aligned}
\log \ttA _{\mathrm{SW}} -  \log \ttA_{\text{loc.}} &= q+ \frac{q^2}{2} + \cdots  ~, \\ 
\log \ttB _{\mathrm{SW}} -  \log \ttB_{\text{loc.}} &= \frac32 q + \frac34 q^2 + \cdots .
\end{aligned}
\end{equation}
A discrepancy of this kind exists for all conformal gauge theories ---
we will see another example of it in the following section --- and we
will discuss this discrepancy in detail in \Cref{u1factor}.

\subsubsection{SU(3)}

The Seiberg-Witten curve in this case is given by
\begin{equation}
Y^2 = \left(Z^3- \tilde{U}_{1} Z^2+ \tilde{U}_{2} Z-1\right)^2 - \frac{4q}{(1+q)^2}\prod_{\ell=1}^6(Z-M_\ell)~.
\end{equation}
The parameters $U_1$ and $U_2$ are determined by the relations
\cref{ccrelationU1} and \cref{ccrelationU2}, with the chiral
correlators $V_1$ and $V_2$ determined from localisation to be
\begin{equation}
\begin{aligned}
  V_1 &=A_1+A_2+\frac{1}{A_1 A_2}+ \frac{q}{A_1A_2} \Bigg[ \frac{
    A_1^2A_2^3\prod_{\ell=1}^6(A_1-M_\ell) }{\left(A_1-A_2\right){}^2
    \left(A_1^2 A_2-1\right){}^2} \\
  &\quad \quad \quad +\frac{A_2^2
    A_1^3\prod_{\ell=1}^6(A_2-M_\ell) }{\left(A_1-A_2\right){}^2 \left(A_1
      A_2^2-1\right){}^2} +\frac{\prod_{\ell=1}^6 (A_1A_2 M_\ell -1)}{\left(A_1^2 A_2-1\right){}^2 \left(A_1 A_2^2-1\right){}^2}\Bigg]+ \cdots \ , \\
  V_2 &=A_1^2+A_2^2+\frac{1}{A_1^2 A_2^2}+ \frac{4q}{A_1^2A_2^2}
  \Bigg[ \frac{A_1^4 A_2^4\prod_{\ell=1}^6(A_1-M_\ell)
  }{\left(A_1-A_2\right){}^2 \left(A_1^2 A_2-1\right){}^2}
  \\
  &\quad \quad \quad +\frac{A_1^4 A_2^4\prod_{\ell=1}^6(A_2-M_\ell) }{\left(A_1-A_2\right){}^2
    \left(A_1 A_2^2-1\right){}^2} +\frac{\prod_{\ell=1}^6 (A_1A_2 M_\ell -1)}{\left(A_1^2 A_2-1\right){}^2
    \left(A_1 A_2^2-1\right){}^2}\Bigg]+ \cdots  \ .
\end{aligned}
\end{equation}
Given these, the functions $\log\ttA$ and $\log\ttB$ can
be calculated from the curve. The expressions are not particularly
illuminating, so we will not reproduce them here. For our purposes,
however, it suffices to observe that just as in the SU$(2)$ case, the
functions $ \log \texttt{A} $ and $ \log \texttt{B} $ computed via (i)
the Seiberg-Witten geometry, and (ii) via localisation match exactly
up to purely $q$-dependent terms. We find that
\begin{equation}
  \label{eq:discrepancy-su3}
\begin{aligned}
  \log \ttA _{\mathrm{SW}} - \log \texttt{A}_{\mathrm{loc.}} &=\frac{3}{2}q+\frac{3}{4}q^2 + \cdots ~, \\ 
  \log \ttB _{\mathrm{SW}} - \log \texttt{B}_{\mathrm{loc.}} &= \frac94q+ \frac{9}{8}q^2+\cdots ~.
\end{aligned}
\end{equation}
.

\section{The Four Dimensional Limit and the U(1) Factor}
\label{u1factor}
For the SU$(N)$ gauge theory with $N_f=2N$ flavours, we observed a
mismatch (see \cref{eq:discrepancy-su2} for $ N=2 $ and
\cref{eq:discrepancy-su3} for $ N=3 $) in the result for $\log\ttA$
and $\log\ttB$ computed from the Seiberg-Witten curve and that
computed from localisation. It is important to emphasise that
\emph{all} the non-trivial dynamical dependence on the Coulomb vevs
and the masses matches, and the mismatch between the two can be
parametrized by a constant (on the Coulomb branch) function of the
instanton counting parameter or, equivalently, the gauge coupling.

\subsection{The Four Dimensional Limit}

In this section we attempt to give a uniform characterization (for any
rank) of the mismatch between the results of the curve and
localisation results for $\log\ttA$ and $\log\ttB$.  However the
calculations in the conformal five-dimensional theory turn out to be
progressively difficult from a computational point of view for the
higher-rank cases. We therefore take the $\beta\rightarrow 0$ limit
and study the four-dimensional SU$(N)$ gauge theory with $N_f=2N$
fundamental matter. The mismatch between the results from the curve
and localisation is identical; this is not surprising as the mismatch
is purely $q$-dependent and independent of the circle radius.

The ingredients needed to carry out this program are once again the
Seiberg-Witten geometry and the resolvent of the four-dimensional
gauge theory, apart from the expectation values for the $ 1 $-point
functions of single trace operators on the Coulomb branch. All of
these are already available in the literature, so we begin with a
brief review of the relevant results. The Seiberg-Witten curve is
given by
\begin{equation}
  y^2 = \hat{P}_N(z)^2 - \frac{4q}{(1+q)^2} \,\hat{B} _{2N}(z) \ ,
\end{equation}
where 
\begin{equation}
  \begin{aligned}
    \hat{P}_N(z) &= z^N + \tilde{u}_{2} z^{N-2} + \cdots + (-1)^N\, \tilde{u}_{N} \ , \\
  \text{and}\quad  \hat{B} _{2N}(z) &= z ^{2N} + s _{2} z ^{N-2} + \cdots + (-1)^{2N}s _{2N} \ .
  \end{aligned}
\end{equation}
As was the case with the curve of the five-dimensional theory, we
define
\begin{equation}
\tilde{u}_k = \left(\frac{1-q}{1+q}\right) u _{k} \ .
\end{equation}
The $ u _{k} $ are the gauge-invariant coordinates on the
four-dimensional Coulomb moduli space, and the $ s _{k} $ are
elementary symmetric polynomials in the fundamental masses. Note that
we have imposed $ \mathrm{SU}(2N) $ flavour symmetry at the level of
the curve, which is the condition $ s _{1} = 0 $. As before, the key
ingredient in the calculation is the resolvent of the gauge theory,
which is given by \cite{Billo:2012st,Ashok:2019rwa}
\begin{equation}
\left\langle \mathrm{Tr} \, \log\frac{z-\Phi}{\mu} \right\rangle = \log \left( (1+q) \frac{\hat{P}_N+y}{2} \right)~. 
\end{equation}
By doing a large-$z$ expansion and comparing coefficients, we read off
the $u$ vs.~$v$ relations, where the $v$'s refer to the single trace
operators $\langle \Tr \Phi^{\ell}\rangle$. We refer the reader to
\cite{Ashok:2019rwa} for details and present only the first couple of
relations:\footnote{The $u_k$ defined in the present work are all
  uniformly rescaled by a factor of $\left(\frac{1-q}{1+q}\right)$
  with respect to the $u_k$ defined in \cite{Ashok:2019rwa}.}
\begin{align}
u_2 &= -\frac12 \left\langle  \mathrm{Tr}\, \Phi^2\right\rangle  + \frac{q}{1-q}\, s_2 \ , \\
u_3 &= +\frac13\, \left\langle  \mathrm{Tr}\, \Phi^3\right\rangle  + \frac{q}{1-q}\, s_3 \ .
\end{align}

We can now repeat what was done in the five-dimensional case, and 
compute the $\log\ttA$ and $\log\ttB$ functions from localisation,
along with the vevs of the single trace operators needed to compute
the $u_k$ on the Coulomb branch. The localisation calculations are
done for a U$(N)$ gauge theory with $N_f=2N$ fundamental flavours, on
which the constraints $\sum_{i}a_i =0$ and $\sum_{\ell}m_\ell=0$ are
subsequently imposed.

For the results from the curve, we calculate
\begin{equation}
  {\texttt A} = \widetilde\alpha~ \text{det} \left(\frac{\mathrm{d} u_i}{\mathrm{d} a_j}\right)^{1/2} \quad \mathrm{and} \quad  {\texttt B} = \widetilde\beta~ \left[\Delta_{\text{phys.}}\left( u _{i} \right)\right]^{1/8}~,
  \end{equation}
  where, for the four dimensional theory we define the physical
  discriminant to be
\begin{equation}
\Delta_{\text{phys.}}(u_i) = \frac{\Delta_{\text{SW}}(u_i)}{\text{Coeff$_{u_N^{4N-2}}(\Delta_{\text{SW}})$}}~,
\end{equation}
where $4N-2$ is the highest degree of $u_N$ in the discriminant of the
four dimensional curve.

We once again isolate the discrepancies between curve- and
localisation-based computations of the effective gravitational
couplings $ \log \texttt{A} $ and $ \log \texttt{B} $, which we will
call $ \log \delta _{\texttt{A}} $ and $ \log \delta _{\texttt{B}} $
respectively. These results are compiled for low-rank gauge groups in
Table \ref{tab:the-only-table}.

\begin{table}[ht]
\centering
\begin{tabular}[t]{c>{\centering}p{0.3\linewidth}>{\centering\arraybackslash}p{0.3\linewidth}}
  \toprule
  $ N $ & $ \log \delta _{\texttt{A}} $ & $ \log \delta _{\texttt{B}} $ \\
  \midrule
  $ 2 $ & $q+ \frac{1}{2}q^2+\frac{1}{3}q^3+\frac{1}{4 }q^4 + \cdots  $ & $ \frac32q+\frac34q^2 +\frac{1}{2}q^3+ \frac{3}{8}q^4+\cdots $ \\
  $ 3 $ & $\frac{3}{2}q+\frac{3}{4}q^2 + \cdots  $ & $\frac94q + \frac{9}{8}q^2 + \cdots $\\
  $ 4 $ & $ 2q + q ^{2} + \cdots $ &  $ 3q + \frac32 q ^{2} + \cdots  $\\
 $ 5 $ & $ \frac{5}{2}q +\frac{5}{4} q^{2} + \cdots $ &  $ \frac{15}{2}q + \frac{15}{8} q ^{2} + \cdots  $\\
 $ 6 $ & $ 3q +\frac{3}{2} q^{2} + \cdots $ &  $ \frac92q + \frac94 q ^{2} + \cdots  $\\
   \bottomrule
\end{tabular}
\caption{The differences between curve- and localisation-based
  computations of effective gravitational couplings for gauge groups
  of low rank. These discrepancies are computed in the
  four-dimensional theory with gauge group SU$(N)$ and $N_f=2N$
  fundamental flavours.}
    \label{tab:the-only-table}
\end{table}%
Note that the $ 1 $- and $ 2 $-instanton contributions in the first
two rows are identical to the results obtained in the five-dimensional
case in the previous section.  Also, these results can just as easily
be obtained in the massless theory, since at this order in the
expansion in terms of the $ \Omega $ deformation parameters, on
dimensional grounds there can be no mass-dependent terms. The
$ 4 $-instanton result for the $ \mathrm{SU}(2) $ $N_f=4$ theory and
all the results for the $ N=4, 5, 6 $ cases were obtained in the limit
in which we take the flavours to be massless.

Based on our study of these discrepancies for gauge groups of low rank
and up to a few instantons, we conjecture the following:
\begin{equation}
  \label{eq:an-bn-proposals}
\begin{aligned}
  \log \delta _{\texttt{A}} &= -\frac{N}{2} \log \left( 1-q \right)  \quad \mathrm{and} \quad  \log \delta _{\texttt{B}} &=-\frac{3N}{2} \log \left( 1-q \right)  \ .
\end{aligned}
\end{equation}
Although we do not have a proof of the above conjectures from first
principles, these formulas agree on all test cases we studied and
exactly capture the discrepancies as a function of the gauge coupling
to all orders and the rank of the gauge group.

\subsection{The U(1) Factor}

Earlier efforts to characterise such mismatches (between curve- and
localisation-based computations) of $\ttA$ and $\ttB$ for rank-$ 1 $
theories with flavour \cite{Manschot:2019pog} and the higher-rank
${\mathcal N}=2^\star$ theories \cite{John:2022yql} successfully
attributed it to the contribution of a $ \mathrm{U}(1) $ factor, which
has its origins in the AGT correspondence \cite{Alday:2009aq}. In that
context, factoring out the contribution of the $ \mathrm{U}(1) $
factor was crucial in order to match Liouville conformal blocks with
the instanton partition function of conformal quiver gauge theories
with $\mathrm{SU}(2)$ gauge groups. Subsequent work
\cite{Wyllard:2009hg} found that a similar factorization was required
to match the instanton partition function of higher-rank gauge
theories with the Toda conformal blocks.

The origin of the U$(1)$ factor can be understood by recalling that
the Nekrasov integrand is naturally defined for $ \mathrm{U}(N) $
theories, while the curve we have worked with is for the
$ \mathrm{SU}(N) $ theories. Therefore, in order to make comparisons
between the results of curve- and localisation-based computations, one
must take into account the global $ \mathrm{U}(1) $ factor. In terms
of partition functions, we have the decomposition
\begin{equation}
  Z _{\mathrm{U}(N)} = Z _{\mathrm{U}(1)}^{(N)} Z _{\mathrm{SU}(N)} \ .
\end{equation}
It is easily checked that the discrepancies in
\cref{eq:an-bn-proposals} can be accounted for by considering a $
\mathrm{U}(1) $ factor of the form
\begin{equation}
\label{4de0discrepancy}
\log\, Z_{\mathrm{U}(1)}^{(N)} = \cdots + \frac{N}{\epsilon_1\epsilon_2} \left(\frac{\epsilon_1+\epsilon_2}{2}\right)^{2} \log \left(1-q\right) + \cdots \ ,
\end{equation}
where the $ \cdots $ indicate other terms in a small deformation
expansion. Based on our study of just the effective gravitational
couplings, this is as much as we can say about the $ \mathrm{U}(1) $
factor with absolute certainty. It is, however, a small step to
conjecture the following U$(1)$ factor for the four dimensional
SU$(N)$ gauge theory with $2N$ flavours with masses $m_i$ in the
fundamental representation:
\begin{equation}
Z_{\mathrm{U}(1),4 \mathrm{d} }^{(N)} = \left( 1-q \right)^{\frac{N}{4\epsilon_1\epsilon_2}\Big(\sum_{i=1}^N m_i + \epsilon_1+\epsilon_2\Big)\Big(\sum_{i=N+1}^{2N} m_i + \epsilon_1+\epsilon_2\Big)} \ .
\end{equation}
Firstly, this reduces to the factor proposed in
\cite{Manschot:2019pog} for the case of $N=2$.  Secondly, at linear
order in the $\epsilon_i$-expansion, the contribution to $\ttH$ is
proportional to the sum of all the masses of the fundamental flavours,
which vanishes due to the SU$(2N)$ flavour constraint. At second order
in the $\epsilon_i$-expansion, we recover the factor derived in
\cref{4de0discrepancy} for all $N$. Finally, at leading order,
this introduces a quadratic term in the masses that breaks the Weyl
symmetry acting on the masses. One can proceed, as in
\cite{Manschot:2019pog}, to define a Weyl-invariant prepotential (and
associated Coulomb vev) by subtracting out a constant
(moduli independent) term.

Let us now return to the Kaluza-Klein theories. As we have seen, in
the $\epsilon_i\rightarrow 0$ limit, the discrepancy in the
Kaluza-Klein compactified theory is identical to that of the
four-dimensional theory.  This is essentially because
$\epsilon_i\rightarrow 0$ is identical to the $\beta\rightarrow 0$
limit. It is therefore a simple matter to lift the above formula for
the U$(1)$ partition function to five dimensions. In terms of the
exponentiated $ \Omega $ deformation parameters
$ E _{i} = e^{\beta \epsilon _{i}} $, and mass parameters
$ M_i = e^{\beta m_i} $, we have:
\begin{equation}
\label{proposedu15d}
Z _{\mathrm{U}(1), 5 \mathrm{d} }^{(N)} = \left( 1-q \right)^{ \frac{N}{\left( E _{1}-1 \right)\left(E _{2}-1 \right)} \frac{1}{4}  \left( M_1\cdots M_N E _{1}E _{2} - 1\right)  \left( M_{N+1}\cdots M_{2N} E _{1}E _{2} - 1\right)} \ .
\end{equation}

Taking into consideration the contribution of the above U$(1)$ factor
and the perturbative ($ 1 $-loop) contribution (see Appendix
\ref{Zpert} for details) we find that with the following choices:
\begin{equation}
  \widetilde{\alpha } = (\beta\Lambda)^{-\frac{N(N-1)}{4}} \quad \mathrm{and} \quad \widetilde{\beta } =  (\beta\Lambda)^{-\frac{N(2N-1)}{4}}  \ ,
  \label{betatildeqfn}
\end{equation}
we find perfect agreement between curve and localisation results. As
in the case of the pure gauge theory, the factors $ \widetilde{\alpha
}  $ and $ \widetilde{\beta }  $ are functions of the dimensionless
constant $ \beta \Lambda  $, which ensures that the five-dimensional
results smoothly go over into the four-dimensional results in the
limit of vanishing circle radius.

\section{Summary and Discussion}
\label{Discussion}

Our investigations in this paper have focused on the case of
four-dimensional theories arrived at via Kaluza-Klein reduction of
five-dimensional $ \mathcal{N}=1 $ supersymmetric gauge theories. We
studied both the pure $ \mathrm{SU}(N) $ gauge theories at admissible
Chern-Simons level, and the conformal gauge theories with
$ N _{f} = 2N $ fundamental hypermultiplets. In studying the case of
non-zero Chern-Simons levels and the case of higher-rank gauge groups,
we have extended earlier work \cite{Manschot:2019pog,Closset:2021lhd}
on the study of these effective gravitational couplings. Our analysis
crucially used the resolvent of the gauge theory, which in turn
allowed us to compute quantum corrections to the Coulomb moduli order
by order in the instanton expansion. In all the theories we studied,
we found that the the effective gravitational couplings independently
determined via equivariant localisation match the expectations in
\cref{rels1} arising from considerations of holomorphy, R-symmetry,
etc. up to the constants of the proportionality.

The case of the conformal gauge theories was particularly interesting,
as we found a non-dynamical, purely $ q $-dependent discrepancy. Such
discrepancies are not new in the study of conformal gauge theories ---
equivariant localisation in these cases reproduces the results derived
from the Seiberg-Witten curve, but only up to constants on the Coulomb
branch. We required that these discrepancies in the effective
gravitational couplings be absorbed by an appropriate
$ \mathrm{U}(1) $ factor. Significantly, our rescaling of the Coulomb
moduli in the curve was crucial for this to work. While many
alternative parametrisations of the Coulomb moduli space are
permissible \cite{Argyres:1999ty}, each will leave its imprint in the
choices of constants of proportionality forced upon us to match the
results of curve- and localisation-based computations. We have
presented in this paper the unique choice attributes all
$ k $-instanton discrepancies between curve- and localisation-based
computations to the $ \mathrm{U}(1) $ factor, in line with other
conformal gauge theories. It would be an interesting task to derive
the full five-dimensional $ \mathrm{U}(1) $ factor, perhaps along the
lines of \cite{Nekrasov:2015wsu,Nekrasov:2017gzb,Jeong:2017mfh}.

Another line of investigation that naturally presents itself is the
question of resummation. As is well-known, the constraints from
S-duality on superconformal $ \mathcal{N}=2 $ gauge theories takes the
form of a modular anomaly equation, which can be used to reconstruct
the prepotential
\cite{Minahan:1997if,Billo:2013fi,Billo:2013jba,Das:2020fhs}. It is
natural to ask if the effective gravitational couplings presented in
this paper are similarly constrained, and whether they can be resummed
into modular forms of the relevant S-duality group. We hope to return
to some of these questions in the near future.

\appendix

\section{Localisation Results}
\label{localizationresults}
\subsection{The Instanton Partition Function}
The partition function of an $\mathcal N=1$ $\mathrm{SU}(N)$ gauge
theory with $ N _{f} $ fundamental hypermultiplets on
$\mathbb R^4 \times \mathbb{S}^1$ and computed via supersymmetric
localisation is given by \cite{Nekrasov:2002qd}:
\begin{equation}
\label{instanton_partition}
Z_{\mathrm{inst.}}=1+ \sum_{k=1}^{\infty}\frac{(-q)^k}{k!}\int_C \prod_{\sigma=1}^k
\left( \beta \frac{\mathrm{d} \chi_{\sigma}}{2\pi i}\right) z_k^{\text{gauge}}(\chi_{\sigma})~z_k^{\text{fund}}(\chi_{\sigma})~.
\end{equation}
where the contribution of the vector and hypermultiplets is captured by the integrands
\begin{equation}
  \label{instantonintegrand}
\begin{aligned}
  z_k^{\text{gauge}}(\chi_{\sigma}) &= e^{-\beta \kappa\sum_{\sigma}\chi_{\sigma}} \prod_{\sigma,\tau=1}^k\left[\frac{g (\chi_{\sigma}-\chi_{\tau}+\epsilon_1+\epsilon_2)}{g (\chi_{\sigma}-\chi_{\tau}+\epsilon_1)~g (\chi_{\sigma}-\chi_{\tau}+\epsilon_2)}\right]\prod_{\sigma \ne\tau=1}^k g(\chi_{\sigma}-\chi_{\tau}) \cr
  &\hspace{2cm} \times \prod_{\sigma=1}^k\prod_{u=1}^N \left[\frac{1}{g\left(\chi_{\sigma}-a_u+\frac{\epsilon_1+\epsilon_2}{2}\right)g\left(-\chi_{\sigma}+a_u+\frac{\epsilon_1+\epsilon_2}{2}\right)}\right]~,\\
  z_k^{\text{fund}}(\chi_{\sigma})  &= \prod_{\sigma=1}^k\prod_{\ell=1}^{N_f} g(\chi_\sigma - m_\ell)~.
\end{aligned}
\end{equation}
In the above expressions,
$g(x)=2\sinh{\left(\frac{\beta x}{2}\right)}$, $\beta$ is the
circumference of the $\mathbb{S}^1$, and $\kappa $ is the
five-dimensional Chern-Simons level. The Coulomb branch of the theory
has been parametrized by the vev $a_u$ with $u=1,\ldots,N$ of the
adjoint scalar field $\Phi$ in the vector multiplet. They satisfy the
$\mathrm{SU}(N)$ condition $\sum_{u=1}^N a_u=0$.

We have studied two models in the main text.
\begin{itemize}
\item For the pure gauge theory in which we omit the contribution due
  to fundamental matter, the instanton counting parameter $q$ is
  related to the complexified strong coupling scale $\Lambda$ and the
  radius of the $\mathbb{S}^1$ by the relation:
\begin{equation}
    (\beta\Lambda)^{2N} =(-1)^N q~.
\end{equation}

\item For the case with conformal matter, we set $N_f=2N$, and the
  instanton counting parameter $q$ is related to $ g ^{2} $, the
  parameter that appears in the Seiberg-Witten curve, by \cref{gvsq}.
\end{itemize}

The integral in \cref{instanton_partition} is a (closed) contour
integral over the complex $\chi_{\sigma}$-planes. We briefly review
the contour prescription now. We take the $a_u$ to be real and assign
an imaginary part to the $\Omega$ deformation parameters, such that
\begin{equation}
0 \ll \text{Im}(\epsilon_2) \ll \text{Im}(\epsilon_1) \ll 1 \ .
\end{equation}
With this choice, the poles in the integrand lie either in the
upper-half plane or the lower-half plane of $\chi_\sigma$. As is well
known \cite{Nekrasov:2002qd,Nakajima:2003pg}, the poles are in a
one-to-one correspondence with Young tableaux, such that the total
number of boxes is equal to $k$, the instanton number. We observe that
the locations of the poles is completely independent of the value of
the five-dimensional Chern-Simons coupling. If $(i,j)$ label the row
and column of the Young tableau, the poles are located at
\be
\label{physicalchipoles}
\chi_{\sigma} = a_u \pm (i-1/2)\epsilon_1 \pm (j-1/2)\epsilon_2 + \frac{2\pi i}{\beta}n~.
\ee
We restrict ourselves to just the fundamental domain, with $n=0$, and
we choose the convention in which we select the poles in the
upper-half planes. This, and the choice of contours at $ 1 $-instanton
has been discussed in great detail in \cite{Ashok:2017bld}, to which
we refer the reader for details. This can be suitably generalized to
higher instantons and we have obtained the $ 2 $-instanton results for
several low-rank cases.

Once we have the answer for the partition function (including the
classical and perturbative contributions) we perform an expansion in
small $\epsilon_i$, and find the following:
\begin{equation}
  \lim_{\epsilon_i\rightarrow 0} \left( \epsilon_1 \epsilon_2 \log Z\right) = -\ttF +(\epsilon_1+\epsilon_2)\ttH+\epsilon_1 \epsilon_2 \log \ttA+\frac{\epsilon_1^2+\epsilon_2^2}{3}\log \ttB+ \ldots 
\end{equation}
Here, $\ttF$ is the prepotential of the gauge theory, that governs the
low-energy effective action on the Coulomb branch. In the class of
theories we study we always find that $\ttH=0$. Given that the
instanton partition function itself is dimensionless and that the
$\Omega$ deformation parameters have unit mass dimension, we see that
the prepotential has mass dimension $ 2 $, and the function $\ttH$ has
mass dimension $ 1 $. Our interest in this work will be on the
dimensionless functions $(\ttA, \ttB)$, that govern the gravitational
couplings of the gauge theory when it is placed on an arbitrary
compact $4$-manifold, in this case the $ \Omega  $-deformed $
\mathbb{C}^{2} $. 

\subsection{1-Instanton Results for the Pure Gauge Theory}

The partition function at 1-instanton takes the form: 
\begin{align}
Z_{1\text{-inst.}}&=
-\frac{e^{-\frac{\beta\kappa}{2}(\epsilon_1+\epsilon_2)}}{g(\epsilon_1)g(\epsilon_2)}
\sum_{i=1}^N\frac{1}{A_i^\kappa}\prod_{j\ne i}\frac{A_i A_j}{(A_j-A_i)(A_i e^{\frac{\beta}{2}(\epsilon_1+\epsilon_2)}-A_je^{-\frac{\beta}{2}(\epsilon_1+\epsilon_2)}}
\end{align}
Here and in what follows, we will always restrict to the case that $|\kappa|  < N$. As explained in \cite{Ashok:2017bld}, this ensures that the instanton partition function receives contributions only from from the physical poles at $\chi_1 = a_u + \frac12(\epsilon_1+\epsilon_2)$, and do not include contributions from $\infty$. 
 We now expand the partition function as a series expansion in $(\epsilon_1, \epsilon_2)$ and keep the leading, sub-leading and the sub-sub-leading terms. We use the variables $s= \epsilon_1+\epsilon_2$ and $p =\epsilon_1 \epsilon_2$, and find that up to second order in the $\epsilon_i$, we have 
\begin{multline}
    \left(\epsilon_1 \epsilon_2 Z_{1\text{-inst.}}\right)=-e^{-\frac{\kappa s}{2}}\left(1-\frac{s^2-2p}{24}\right) \sum\limits_{i=1}^N\frac{1}{{A_i}^\kappa}\prod\limits_{j\ne i}\Big[-\frac{A_i A_j}{A_{i j}^2}+\frac{s}{2}\frac{A_i A_j(A_i+A_j)}{{A_{i j}}^3}\\
    -\frac{s^2}{8}\frac{A_i A_j\left(A_i^2+6A_i A_j+A_j^2\right)}{{A_{i j}^4}}+...\Big]~.
\end{multline}
We can now extract the leading, subleading and the sub-subleading terms in the $\epsilon_i$, giving us the $(\ttF, \ttA, \ttB)$ terms: 
\begin{equation}
  \label{eq:b-kappa-zero}
\begin{aligned}
\beta^2\, \ttF_{1\text{-inst.}} &= (-1)^{N-1}\sum_{i=1}^{N}\frac{1}{A_i^\kappa} \prod _{j\neq i} ^{N} \frac{A_iA_j}{\left( A _{i} - A _{j} \right)^{2}} ~,\\[5pt]
    \log \ttB_{1\text{-inst.}}&=\frac{3 (-1)^{N}}{8}\sum\limits_{i=1}^N\frac{1}{A_i^\kappa}\left[\sum\limits_{j\ne i, k\ne i, k\ne j}\frac{A_iA_j(A_i+A_j)}{(A_i-A_j)^3}\frac{A_iA_k(A_i+A_k)}{(A_i-A_k)^3}\prod\limits_{l\ne i,l\ne j,l\ne k}\frac{A_i A_l}{(A_i-A_l)^2}\right.\cr
&\left. +\left(\kappa^2-\frac{1}{3}\right)\prod\limits_{j\ne i}\frac{A_iA_j}{(A_i-A_j)^2}+\sum\limits_{j\ne i}\frac{A_iA_j\left(A_i^2+6A_iA_j+A_j^2\right)}{(A_i-A_j)^4}\prod\limits_{k\ne j,k\ne i}\frac{A_i A_k}{(A_i-A_k)^2}\right.\cr&\left.+2\kappa\sum\limits_{j\ne i}\frac{A_i A_j(A_i+A_j)}{(A_i-A_j)^3}\prod\limits_{k\ne j,k\ne i}\frac{A_iA_k}{(A_i-A_k)^2}\right]~,\\[5pt]
    \log \ttA_{1\text{-inst.}}&=\frac{ (-1)^{N}}{4}\sum\limits_{i=1}^N\frac{1}{A_i^\kappa}\left[\sum\limits_{j\ne i, k\ne i, k\ne j}\frac{A_iA_j(A_i+A_j)}{(A_i-A_j)^3}\frac{A_iA_k(A_i+A_k)}{(A_i-A_k)^3}\prod\limits_{l\ne i,l\ne j,l\ne k}\frac{A_i A_l}{(A_i-A_l)^2}\right.\cr
&\left. +\kappa^2\prod\limits_{j\ne i}\frac{A_iA_j}{(A_i-A_j)^2}+\sum\limits_{j\ne i}\frac{A_iA_j\left(A_i^2+6A_iA_j+A_j^2\right)}{(A_i-A_j)^4}\prod\limits_{k\ne j,k\ne i}\frac{A_i A_k}{(A_i-A_k)^2}\right.\cr&\left.+2\kappa\sum\limits_{j\ne i}\frac{A_i A_j(A_i+A_j)}{(A_i-A_j)^3}\prod\limits_{k\ne j,k\ne i}\frac{A_iA_k}{(A_i-A_k)^2}\right]~.
\end{aligned}
  \end{equation}

\subsection{Expectation Value of Chiral Operators in the Pure Gauge Theory}

The chiral correlators of the five-dimensional gauge theory can also
be computed using localisation \cite{Ashok:2017bld}. We have
\begin{align}
V_{\ell} = \left\langle \Tr e^{\ell \beta\Phi} \right\rangle = \sum_{u=1}^NA_u^{\ell} - \frac{1}{Z_{\mathrm{inst.}}} \sum_{k=1}^{\infty} \frac{(-q)^k}{k!}\int_{\cal C}\left(\prod_{\sigma=1}^k \frac{\beta \,\mathrm{d} \chi_{\sigma}} {2\pi i} \right)~  z_k^{\text{gauge}}(\chi_{\sigma})~ {\cal O}_{\ell}(\chi_{\sigma})~.
\end{align} 
In the formula, $Z_{\mathrm{inst.}}$ is the instanton partition
function defined in \cref{instanton_partition}, $z_k$ is the same
integrand as in \cref{instantonintegrand} and ${\cal O}_{\ell}$ is
given by
\be
{\cal O}_{\ell}(\chi_{\sigma}) = \sum_{\sigma=1}^k  e^{\ell\beta\chi_\sigma}(1-e^{\ell\beta\epsilon_1})(1-e^{\ell\beta\epsilon_2})~.
\ee
Up to $ 1 $-instanton, the chiral correlators $V_{\ell }$ are given by
\cite{Ashok:2017bld}:
 \begin{equation}
 \label{Vellloc}
    V_{\ell }= \sum_{u=1}^N  A_u^{\ell } +\ell ^2(\beta\Lambda)^{2N}\sum_{u=1}^N\left[\frac{A_u^{N-2+\ell -\kappa}}{\prod_{u\ne v}(A_u-A_v)^2}\right]+\mathcal O\left((\beta\Lambda)^{4N}\right) \ ,
 \end{equation}
 where $A_u \equiv e^{\beta a_u}$ and the $\mathrm{SU}(N)$ condition is
 implemented by
 \begin{equation}
     \prod_{u=1}^N A_u =1~.
 \end{equation}
 Similar expressions can be obtained, both for the partition function
 and for the chiral correlators in the case with fundamental
 flavour. But the expressions are not very illuminating and we present
 just the results for the low rank cases in the main text.

\section{The Perturbative Contribution}
\label{Zpert}

The one loop or perturbative contribution to the partition function of
an $\Omega$-deformed gauge theory of the Kaluza-Klein type that we
have been studying has been obtained in the classic work
\cite{Nakajima:2005fg}. The main ingredient here is the
five-dimensional lift of the special function
$\gamma_{\epsilon_1, \epsilon_2}(x|\Lambda)$ that already appeared in
the earlier works \cite{Nekrasov:2003rj, Nakajima:2003uh}. In what
follows we shall briefly review the definitions and then provide the
particular combinations of the special functions of interest to the
gauge theories studied in the main text.

Following \cite{Nakajima:2005fg} we first of all define 
\begin{equation}
\begin{aligned}
  \gamma_{\epsilon_1, \epsilon_2}(x|\beta, \Lambda) &=\frac{1}{2\epsilon_1\epsilon_2}\left(x^2\log\left(\beta\Lambda\right)-\frac{\beta}{6}\Big(x+\frac12(\epsilon_1+\epsilon_2) \Big)^3\right)\\
  &\quad +\sum_{m\ge 0} \frac{c_m}{m!} \beta^{m-2}\text{Li}_{3-m}(e^{-\beta x})~.
\end{aligned}
\end{equation}
Here the $\text{Li}_k$is the $k$th polylogarithm and the $c_m$ are obtained as coefficients in the expansion:
\begin{equation}
\frac{1}{(1-e^{-\beta\epsilon_1})(1-e^{-\beta\epsilon_2}) } = \sum_{m\ge 0}\frac{c_m}{m!} \beta^{m-2}~.
\end{equation}
We list the first few coefficients that will be of relevance in our calculations:
\begin{equation}
c_0 = \frac{1}{\epsilon_1\epsilon_2}~, \quad c_1 = -\frac{\epsilon_1+\epsilon_2}{2\epsilon_1\epsilon_2}~, \quad c_2 = \frac{\epsilon_1^2+3\epsilon_1\epsilon_2+\epsilon_2^2}{6\epsilon_1\epsilon_2}~.
\end{equation}
Then, the relevant function in terms of which the perturbative
contribution will be written is given by
\begin{equation}
\label{gammatilde}
\begin{aligned}
  \widetilde{\gamma}_{\epsilon_1, \epsilon_2}(x|\beta, \Lambda) &=\gamma_{\epsilon_1, \epsilon_2}(x|\beta, \Lambda)+\frac{1}{\epsilon_1\epsilon_2}\left(\frac{\pi^2x}{6\beta}-\frac{\zeta(3)}{\beta^2} \right)\\
  &\quad +\frac{\epsilon_1+\epsilon_2}{2\epsilon_1\epsilon_2}\left(x\log(\beta\Lambda) + \frac{\pi^2}{6\beta} \right) +\frac{\epsilon_1^2+3\epsilon_1\epsilon_2+\epsilon_2^2}{12\epsilon_1\epsilon_2}\log(\beta\Lambda) + \cdots ~.
\end{aligned}
\end{equation}
It is shown in \cite{Nakajima:2005fg} that this function has the
expected four dimensional behaviour in the limit that
$\beta\rightarrow 0$. Our present goal is to find linear combinations
that give the correct logarithmic terms in $\log\ttA$ and $\log\ttB$
for the pure gauge theory and in the case with flavour in the limit
$q\rightarrow 0$. For this purpose we consider the small-$\epsilon_i$
expansion of the function in \cref{gammatilde}, which is valid for
$x>0$:
\begin{align}
\widetilde{\gamma}_{\epsilon_1, \epsilon_2}(x|\beta, \Lambda) =&\phantom{+}\frac{1}{\epsilon_1\epsilon_2}\frac{1}{\beta^2}\left( \text{Li}_3(e^{-\beta x})-\zeta(3) - \frac{\pi^2}{6}\beta x+\frac{\beta^2x^2}{2}\log(\beta\Lambda)-\frac{\beta^3 x^3}{12}\right) \nonumber\\
&+\frac{\epsilon_1+\epsilon_2}{2\epsilon_1\epsilon_2 }\frac{1}{\beta}\left(\frac{\pi^2}{6}-\text{Li}_2(e^{-\beta x})+\beta x\log(\beta\Lambda)-\frac{\beta^2 x^2}{4}\right)\nonumber\\
&-\frac{(\epsilon_1^2+\epsilon^2+3\epsilon_1\epsilon_2)}{12\epsilon_1\epsilon_2}\log\left(\frac{1-e^{-\beta x}}{\beta\Lambda}\right) - \frac{\beta x}{16}(\epsilon_1+\epsilon_2)^2 + \cdots
\end{align}
The linear combination shown below proves to be useful to write down the perturbative contribution for the pure gauge theory; we present the small-$\epsilon _{i}$ expansion:
\begin{align}
\widetilde{\gamma}^{\text{gauge}}_{\epsilon_1, \epsilon_2}(x|\beta, \Lambda) =& -\widetilde{\gamma}_{\epsilon_1, \epsilon_2}(x|\beta, \Lambda) -\widetilde{\gamma}_{\epsilon_1, \epsilon_2}(x-\epsilon_1-\epsilon_2|\beta, \Lambda)\nonumber  \\
=&-\frac{1}{\epsilon_1\epsilon_2}\frac{1}{\beta^2}\left(2\Big(\text{Li}_3(e^{-\beta x})-\zeta(3)\Big) + \frac{\pi^2}{3}\beta x +\beta^2 x^2 \log(\beta\Lambda)-\frac{\beta^3 x^3}{6}\right)\nonumber \\
&+\left(\frac{\epsilon_1^2+\epsilon_2^2}{3\epsilon_1\epsilon_2}+1\right) \frac12 \log\left(\frac{e^{\beta x} -1}{\beta\Lambda}\right) - \frac{\epsilon_1^2+\epsilon_2^2+6\epsilon_1\epsilon_2}{24\epsilon_1\epsilon_2}\beta\, x + \cdots
\end{align}
This combination ensures that $\ttH = 0$, and that at second order in
the expansion, the contribution to both the $\log\ttA$ and $\log\ttB$
terms is identical as far as the log-terms are concerned. The
contribution of the linear term cancels out once we sum over all roots
of the Lie algebra.

The following shifted function in turn allows one to write down the
perturbative contribution from the matter fields in the fundamental
representation:
\begin{align}
\widetilde{\gamma}^{\text{fund.}}_{\epsilon_1, \epsilon_2}(x|\beta, \Lambda) =&\phantom{+}\widetilde{\gamma}_{\epsilon_1, \epsilon_2}\left(x-\frac12(\epsilon_1+\epsilon_2)\bigg\vert\beta, \Lambda\right) \nonumber\\
=&\phantom{+}\frac{1}{\epsilon_1\epsilon_2}\frac{1}{\beta^2}\left(\text{Li}_3(e^{-\beta x})-\zeta(3)+ \frac{\pi^2 }{6} \beta x + \frac{\beta^2x^2}{2} \log(\beta\Lambda)-\frac{\beta^3 x^3}{12}\right)\nonumber\\
&+\frac{(\epsilon_1^2+\epsilon_2^2)}{3\epsilon_1\epsilon_2}\frac18\left[\log\left(\frac{e^{\beta x} -1}{\beta\Lambda}  \right)-\beta x \right] + \cdots
\end{align}
In this case the shift of the argument of the
$\widetilde\gamma$-function ensures that the contribution to both
$\ttH$ as well as $\log\ttA$ vanishes.

The partition function of the $\Omega$-deformed Kaluza-Klein gauge
theory, compactified on a circle of circumference $\beta$, with gauge
group SU$(N)$ and with $N_f$ flavours in the fundamental
representation is given by
\begin{align}
\log Z_{\text{pert.}} = \frac12\sum_{i\ne j} \widetilde{\gamma}^{\text{gauge}}_{\epsilon_1, \epsilon_2}(a_i-a_j|\beta, \Lambda) + \sum_{i=1}^{N}\sum_{\ell =1}^{N_f}\widetilde{\gamma}^{\text{fund.}}_{\epsilon_1, \epsilon_2}(a_i - m_{\ell}|\beta, \Lambda)~,
\end{align}
with the constraints $\sum_{i=1}^N a_i = 0$ and
$\sum_{\ell=1}^{N_f} m_{\ell} = 0$ imposed on the Coulomb v.e.vs and mass
parameters (this also ensures the vanishing of the linear terms in the
combination of special functions that contribute to the one loop contribution for the fundamental representation). 
We note that in the four-dimensional
limit, the same combinations of the special functions with shifted
arguments occur in the $ 1 $-loop contribution proposed in
\cite{Manschot:2019pog}.

\bibliography{refs-kk-theories.bib}
\bibliographystyle{JHEP}

\end{document}